\documentclass[aps,prl,twocolumn,showpacs,floatfix]{revtex4}

\usepackage{dcolumn}
\usepackage{amssymb}
\usepackage{graphicx}

\begin{document}

\newcommand \be {\begin{equation}}
\newcommand \ee {\end{equation}}
\newcommand \bea {\begin{eqnarray}}
\newcommand \eea {\end{eqnarray}}
\newcommand \nn {\nonumber}
\newcommand \la {\langle}
\newcommand \rl {\rangle_L}

\title{Subdiffusion and dynamical heterogeneities in a lattice glass
model}
\author{Eric Bertin,$^{1,2}$ Jean-Philippe Bouchaud,$^2$ and Fran\c{c}ois
Lequeux$^3$}
\affiliation{
$^1$ Department of Theoretical Physics, University of Geneva,
CH-1211 Geneva 4, Switzerland\\
$^2$ SPEC, CEA Saclay, F-91191 Gif-sur-Yvette Cedex, France\\
$^3$ ESPCI, 10 rue Vauquelin, F-75005 Paris, France}

\date{\today}

\begin{abstract}
We study a kinetically constrained lattice glass model in which continuous
local densities are randomly redistributed on neighbouring sites
with a kinetic constraint that inhibits the process at high densities, and 
a random bias accounting for attractive or repulsive interactions.
The full steady-state distribution can be computed exactly in any space
dimension $d$. Dynamical heterogeneities are characterized by a length scale
that diverges when approaching the critical density. 
The glassy dynamics of the model can be described as
a reaction-diffusion process for the mobile regions.
The motion of mobile regions is found to be subdiffusive, for a 
large range of parameters, due to a self-induced trapping mechanism.
\end{abstract}


\maketitle

One of the most important features of glassy dynamics is its heterogeneous
character, that is the coexistence of slowly and rapidly evolving regions,
with a characteristic size significantly larger than the molecular scale.
Such dynamical heterogeneities, which are often due to a jamming 
phenomenon (like for instance in colloids or in granular materials),
have been observed recently both in experimental systems
\cite{Sillescu,Ediger,Weeks}
and in numerical simulations \cite{Heuer,Perera,Donati}.
In order to model these effects,
a fruitful path, which has attracted considerable attention 
in recent years, is to introduce ``Kinetically Constrained Models''
({\sc kcm}s) with a very 
simple (usually one body) hamiltonian. Steric constraints are taken into account through
kinetic rules that forbid some transitions between microscopic states
\cite{RS}.
The study of these {\sc kcm}s has emphasized the r\^ole played in the
relaxation process by rare and localized regions
(often called mobility excitations or defects)
that are not completely blocked by
the constraints \cite{JStatPhys,GC,BerthierJCP03}. These mobility excitations
diffuse throughout the system, eventually leading to full decorrelation.
This simple relaxation mechanism suggests a somewhat universal dynamical behavior, 
as advocated in \cite{Whitelam}. Whether or not this picture applies to
all glasses, it relies on the assumption that mobility
excitations follow a purely diffusive motion, which is not justified on general grounds, as we shall demonstrate below.

In this Letter, we consider a new {\sc kcm} in which the local variable
is a continuous density, rather than a discrete variable as in most studied {\sc kcm}s.
The interest of the present model, compared to previously studied {\sc kcm}s, is two-fold. First, 
the statics of the model is non trivial, and can be characterized exactly. The stationary N-body distribution
turns out to be factorizable for all values of a parameter that describes the (repulsive or attractive) 
interaction between particles, which has no counterpart in other {\sc kcm}s.
Second, the presence of continuous local densities leads to an interesting
dynamical behaviour; the motion of mobility excitations is found to be subdiffusive for a large parameter range,
due to a self-induced trapping mechanism (i.e., not introduced by hand in the model). Accordingly, dynamical
heterogeneities with a rather rich spatial structure are observed. The model is defined on a lattice of arbitrary 
dimension $d$ with $N$ sites. In each cell centred on the lattice site $i$, we define $\rho_i$ as the density of particles. 
The dynamics, aimed at describing density fluctuations, corresponds to a local redistribution of particles across
the links of the lattice. At each time step $\Delta t = \tau_0/N$ ($\tau_0$ is a microscopic time scale),
two neighboring sites $(j,k)$ are chosen at random, and
$\rho_j$ and $\rho_k$ are redistributed to become $\rho_j'$ and $\rho_k'$:
\be
\rho_j' = q (\rho_j+\rho_k), \quad
\rho_k' = (1-q) (\rho_j+\rho_k), \quad 0<q<1.
\ee
Note that the mass $\rho_j+\rho_k$ is exactly conserved at each step. The fraction $q$ is a random variable,
chosen independently both in space and in time, with distribution $\psi(q)$ such that $\psi(1-q)=\psi(q)$,
that plays the r\^ole of internal noise in the model. We want to model the fact that a locally 
dense packing is blocked unless some low density cell is present in its vicinity.
A simple kinetic constraint is to allow redistribution only when $\rho_j+\rho_k < 2 \rho_{th}$; 
in the following, we set $\rho_{th}=1$.
Thus for large densities, the system is no longer able to reorganize locally. 
One can expect that if the average density is high, the dynamics
slows down dramatically and exhibits glassy behaviour.
Note that if initially $\rho_i^0<2$ for all $i$, the evolution rules forbid
any density $\rho_i>2$ at later times.

In order to find an exact solution for the stationary state, we choose a beta distribution
$\psi(q) = \Gamma(2\mu)/\Gamma(\mu)^2 [q(1-q)]^{\mu-1}$. The case $\mu=1$ corresponds to a uniform 
redistribution and may be thought of as non interacting particles (except from hard-core repulsion). The case 
$\mu < 1$ favors $q$ close to zero or to one, and can be interpreted as an effective attraction. Conversely, $\mu > 1$ favors the
maximal mixing value $q=1/2$, and mimics repulsive interactions and suppressed density fluctuations.
Note that a model similar to (but different from) the present model has been studied numerically
in the specific case $\mu=1$ \cite{Lequeux}. From the Master equation describing the model, one 
sees that a non trivial form of detailed balance holds \cite{long},
leading to the exact stationary $N$-body distribution:
\be \label{eq-PNst}
P_{st}(\{\rho_i\}) = \frac{1}{Z_N} \prod_{i=1}^N
[\rho_i^{\mu-1} \theta(2-\rho_i)] \;
\delta \left(\sum_{i=1}^N \rho_i - N\overline{\rho} \right),
\ee
where $\overline{\rho}$ is the average density. The above explicit solution is one of the central results
of this letter. It shows that the stationary distribution is generally not uniform among all available states.
This must be contrasted with the Edwards prescription, often used for generic jamming
problems, which only holds when $\mu=1$. Note that the above steady-state distribution is obtained 
in the long time limit only if, as noted above, all the initial densities
$\{\rho_i^0\}$ are less than $2$ and at least some links initially satisfy $\rho_j^0+\rho_k^0<2$ (otherwise no
redistribution can occur at all). From Eq.~(\ref{eq-PNst}), the `canonical' distribution $P_{can}(\{\rho_i\})$,
describing a subsystem with $K$ sites, can be derived in the limit $1 \ll K \ll N$ \cite{long}:
\be \label{dist-can}
P_{can}(\{\rho_i\}) =  \frac{1}{Z_K^{can}} \, \prod_{i=1}^K [\rho_i^{\mu-1} \theta(2-\rho_i) \, e^{\beta \rho_i}],
\ee
with $\beta=-N^{-1} \partial \ln Z_N/\partial \overline{\rho}$.
The one-site distribution is thus given by
$p(\rho) = c\, \rho^{\mu-1} e^{\beta \rho}$, with $0<\rho<2$
and $Z_K^{can}=c^{-K}$.
In order to characterize more quantitatively
the glassy properties of this model, we compute the
fraction $\eta$ of mobile links, defined as
links $(j,k)$ such that $\rho_j+\rho_k<2$.
In the `canonical' steady state, $\eta$ is
computed as: $\eta =
\int_0^2 d\rho_1 \int_0^2 d\rho_2 \, p(\rho_1) \, p(\rho_2) \,
\theta(2-\rho_1-\rho_2)$ and can be evaluated numerically.
In the limit $\overline{\rho} \to 0$, the kinetic
constraint does not play any r\^ole and $\eta \to 1$. In the more interesting
limit $\overline{\rho} \to 2$, the density of mobile links develops an
essential singularity:
\be \label{eta-stat}
\eta \simeq \frac{2\, \Gamma(\mu)^2}{(2-\overline{\rho}) \, \Gamma(2\mu)}
\, e^{-2/(2-\overline{\rho})}.
\ee
Thus the fraction of mobile links decreases very fast for $\overline{\rho}
\to 2$, but remains finite for any average density $\overline{\rho}<2$,
suggesting that the critical density is $\overline{\rho}_c=2$,
as will be confirmed in $d=1$ below.
This situation is indeed reminiscent of what
happens in the Kob-Andersen model \cite{KobAndersen} in $d=2$, in which
the diffusion coefficient $D$
goes to $0$ with the particle density $\rho$ as: $\ln D \sim (1-\rho)^{-1}$
\cite{Biroli}.

\begin{figure}[t]
\centering\includegraphics[width=8cm,clip]{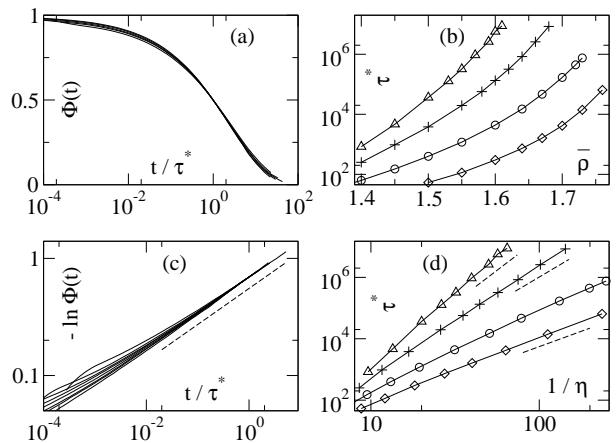}
\caption{\sl (a) Time correlation $\Phi(t)$ versus $t/\tau^*$
for $\overline{\rho}=1.60$ to $1.72$ ($\mu=1$).
(b) $\tau^*$ as a function of $\overline{\rho}$, for $\mu=0.3$ ($\diamond$),
$1$ ($\circ$), $2$ ($+$) and $3$ ($\vartriangle$).
(c) Stretched exponential behavior of $\Phi(t)$ with $\mu=2$, and
exponent $\nu$ determined from $r^2(t)$ (dashed).
(d) Dynamic scaling $\tau^*$ versus $\ell = 1/\eta(\overline{\rho},\mu)$
(same symbols as (b)); dashed: exponent $z=1/\nu$ ($z=2$ for $\mu=0.3$).
}
\label{fig-cortemp}
\end{figure}

All the above analytical results concerning static quantities are valid
in arbitrary dimension $d$. In the following, we present detailed numerical
simulations of dynamical quantities in dimension $d=1$, and discuss briefly
the case $d=2$, where dynamical quantities appear to be more complex.
The relaxation properties of the model can be quantified by
introducing on each site $i$ a function $\phi_i(t)$,
that we choose for simplicity to be the local persistence: $\phi_i(t)=1$
if $\rho_i$ has never changed in the time interval
$[0,t]$, and $\phi_i(t)=0$ otherwise. One can then introduce a global
correlation function $\Phi(t) = [ \langle \phi_i(t)\rangle ]$,
where $\langle \ldots \rangle$ and $[ \ldots ]$ denote averages over
the sites and the noise respectively.
Defining the characteristic decay time $\tau^*$ through
$\Phi(\tau^*)\equiv 1/2$, one can rescale the data by plotting $\Phi(t)$
against $t/\tau^*$ [Fig.~\ref{fig-cortemp}(a)].
The relaxation time $\tau^*$ is plotted as a function of $\overline{\rho}$
for different $\mu$ in Fig.~\ref{fig-cortemp}(b). Interestingly, $\Phi(t)$ behaves 
as a stretched exponential:
$-\ln \Phi(t) \sim (t/\tau^*)^\gamma$, as typical in glassy systems [Fig.~\ref{fig-cortemp}(c)].
The exponent $\gamma$ matches perfectly the conjecture $\gamma=\nu$, where $\nu$ is defined from
the subdiffusion of mobility defects: $r^2(t) \sim t^{2\nu}$ -- see below.

\begin{figure}[t]
\centering\includegraphics[width=7.5cm,clip]{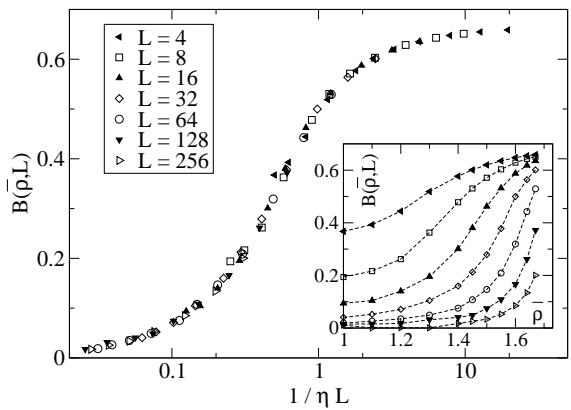}
\caption{\sl Inset: plot of $B(\overline{\rho},L)$ as a function of
$\overline{\rho}$ for different values of $L$ ($\mu=1$). The curves do not
cross, except presumably for $\overline{\rho}=2$. Main plot:
$B(\overline{\rho},L)$ plotted as a function of the rescaled variable
$1/(\eta L)$. The collapse is very good, showing that the characteristic
length scale $\ell$ is proportional to the inverse of the concentration
$\eta$ of mobile links.}
\label{fig-BTL1d}
\end{figure}

Can a non trivial cooperative length scale be associated
with the increased glassiness as $\overline{\rho} \to 2^-$?
Clearly, since the stationary distribution is factorized, no static correlation length can grow in this regime. 
Thus such a length scale can only appear in
dynamical quantities, such as four-point correlation functions that have
been studied recently to describe dynamical heterogeneities
\cite{Glotzer,BerthierPRL03,BerthierJCP03,BiroliBouchaud04}.
In physical terms, these dynamical heterogeneities can be interpreted as
the appearance of `fast' and `slow' regions, with a typical size that grows as the glass
transition is approached. More precisely, one can introduce on each site $i$ a local variable
$\phi_i^* \equiv \phi_i(\tau^*)$. This allows one to define slow sites, such that
$\phi_i^*=1$, whereas fast sites have $\phi_i^*=0$.  This particular choice of $\phi_i^*$ ensures that 
fast and slow regions occupy equal volumes.
To identify quantitatively the characteristic length scale $\ell$ of dynamical
heterogeneities, we study
the fourth (Binder) cumulant $B(\overline{\rho},L)$ of the random variable 
$\omega \equiv L^{-d} \sum_i \phi_i^*$, where the sum is over all sites of a system of size $L$ \cite{BerthierPRL03}. 
This quantity measures the `non-Gaussianity' of $\omega$; it is zero for large systems, 
$L \gg \ell$, and equal to a certain constant ($2/3$ with Binder's
normalization) for $L \ll \ell$.
Using finite size scaling arguments, one expects  $B(\overline{\rho},L)$
to scale as a function of $\ell/L$. We indeed find in the present model that
all data rescale perfectly when plotted as a function of $\eta L$ [Fig.~\ref{fig-BTL1d}].
Thus $\ell$ is simply the average distance $1/\eta$ between mobile links.
It is then natural to look for a scaling relation $\tau^* \sim \ell^z$ between the relaxation time 
and the cooperative length, where $z$ is a dynamical critical exponent; $\tau^*$ is plotted
as a function of $\ell = 1/\eta$ on a log-log scale in Fig.~\ref{fig-cortemp}(d).
Interestingly, the relation between $\tau^*$ and $\ell$ is found to depend strongly on $\mu$; for $\mu=0.3$, 
$2$ and $3$, the data is well fitted by a power-law for $\ell \gg 1$, whereas for $\mu=1$, 
some systematic curvature appears.

As mentioned in the introduction, the high density dynamics can be described 
in terms of reaction-diffusion of mobility excitations \cite{JStatPhys}. When a redistribution occurs on a
given link, the `state' of the link (mobile or not) cannot change, due to mass conservation.
However, neighboring links {\it can} change state since the mass associated to these links is not
conserved by this redistribution. So it is possible to create or destroy a mobile link when it shares a site
with another mobile link. Denoting mobile links by $A$ and immobile links by $\emptyset$, one can write 
schematically these processes as $(A, \emptyset) \to (A, A)$ and $(A,A) \to
(\emptyset,A)$, whereas the simple annihilation process $A \to \emptyset$ is
forbidden by the conservation rule. Moreover, a chain of two such transitions is equivalent
to the motion of a `defect' $A$. At high density $\overline{\rho}$, i.e.~at
low concentration of mobility, branching and annihilation of mobility
excitations become rare, and mobility motion is the dominant
relaxation mechanism (for a related discussion, see \cite{Whitelam}).

\begin{figure}[t]
\centering\includegraphics[width=8cm,clip]{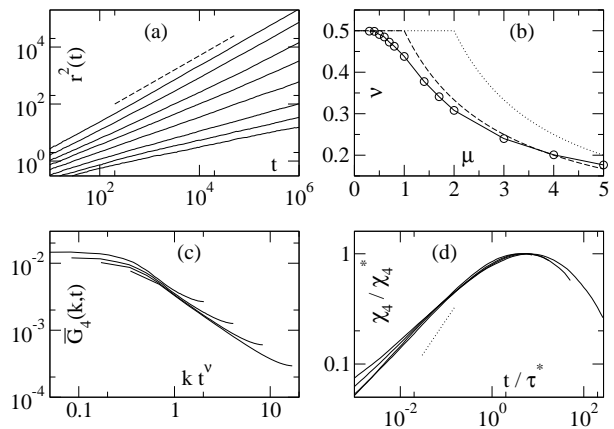}
\caption{\sl (a) Mean-square displacement $r^2(t)$ for $\mu=0.3$, $0.6$, $1$,
$1.4$, $2$, $3$, $4$ and $5$ (top to bottom) and $\overline{\rho}=1.75$;
dashed: slope 1.
(b) Exponent $\nu$ defined by $r^2(t) \sim t^{2\nu}$ ($\circ$), annealed
(dotted) and quenched (dashed) barrier models predictions.
(c) Rescaled correlation $\overline{G}_4(k,t)$ against $k\,t^{\nu}$
($\overline{\rho}=1.60$, $\mu=2$).
(d) Rescaled susceptibility $\chi_4(t)/\chi_4^*$ versus $t/\tau^*$ for
$\overline{\rho}=1.50$,
$1.55$, $1.58$ and $1.60$ ($\mu=2$); dots: slope $2\nu$.
}
\label{fig-subdif}
\end{figure}

If the motion of mobility excitations was purely diffusive, as
$r^2(t) \sim D t$, with a diffusion constant $D$ that remains non zero
as $\overline{\rho} \to 2$, then $\tau^*$ would scale with $\ell$ as
$\tau^* \sim \ell^2$, i.e.~$z=2$. This scaling law is not compatible with the data shown in
Fig.~\ref{fig-cortemp}(d), at least when $\mu \ge 1$.
Such a discrepancy could come from a critical dependence of the diffusion constant $D$
on $\overline{\rho}$. This mechanism is indeed at play in the Fredrickson-Andersen (FA) model in $d=1$, 
leading to $z=3$ \cite{BerthierJCP03}. However, the mechanism operating in the present model is different,
and related to a genuine subdiffusive motion of individual defects. We show in Fig.~\ref{fig-subdif}(a) the mean-square
displacement $r^2(t)$ of mobility excitations for several values of
$\mu$. The motion of mobility is found to be subdiffusive for a whole range
of parameter $\mu$, which we estimate to be $\mu \gtrsim 1$ (finite
time effects induce some uncertainty on this threshold value).
The exponent $\nu$ characterizing the asymptotic power law regime
$r^2(t) \sim \tilde{D}(\overline{\rho}) \, t^{2\nu}$
is shown in Fig.~\ref{fig-subdif}(b) where $\nu$ is independent of $\overline{\rho}$ at high enough density.
This relation suggests $\ell^2 \sim \tilde{D}(\overline{\rho}) \tau^{*2\nu}$;
neglecting the dependence of $\tilde{D}$ on $\overline{\rho}$ leads to $z=1/\nu$.
Using the measured values of $\nu$, one can compare this prediction with the direct evaluation of $z$
on Fig.~\ref{fig-cortemp}(d) for $\mu=2$ and $3$ (dashed lines).
The agreement is quite good, although some discrepancies appear, presumably
due to the density dependence of $\tilde{D}$. For $\mu=0.3$, one recovers the standard exponent $z=2$.
The subdiffusive motion of mobility also accounts very well for the
stretched exponential behavior of $\Phi(t)$. At short time
($t \ll \tau^*$), the correlation should behave as $1-\Phi(t) \sim r(t)/\ell$, i.e.~as $t^{\nu}$.
Thus $\Phi(t)$ is well approximated by a
stretched exponential with exponent $\nu$, which indeed matches perfectly the
numerical data even for $t \sim \tau^*$ -- see Fig.~\ref{fig-cortemp}(c).
Interestingly, the dynamics of mobility can be mapped onto a one-dimensional
barrier model \cite{BG}, since the processes $(\emptyset,A,\emptyset) \to
(\emptyset,A,A)$ and $(\emptyset,A,\emptyset) \to (A,A,\emptyset)$
involve waiting times which become broadly distributed when
$\overline{\rho} \to 2$ \cite{long}.
The absence of quenched disorder suggests to consider an annealed model,
where each random barrier is drawn anew after a jump.
This would lead to $\nu=1/\mu$ for $\mu>2$ \cite{BG,long},
which does not match the numerical data [Fig.~\ref{fig-subdif}(b)].
But as the environment of a mobile link is frozen for times of the order of
$\tau^*$, the quenched barrier model might be more appropriate.
Indeed, the corresponding prediction
$\nu=1/(1+\mu)$ for $\mu>1$ is in rather good agreement with our numerics.
Therefore, the existence of slow regions induces non trivial correlations
that mimic the presence of quenched disorder,
which is actually `self-generated'.

\begin{figure}[t]
\includegraphics[width=2.5cm]{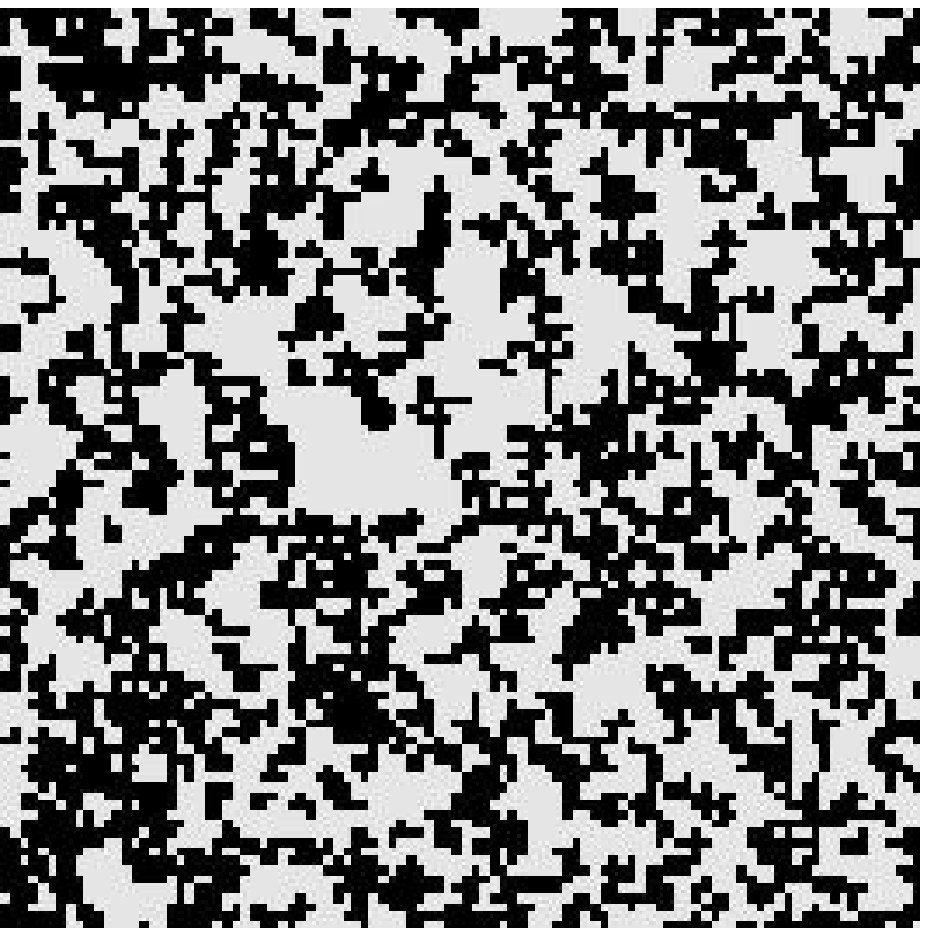}
\hfill
\includegraphics[width=2.5cm]{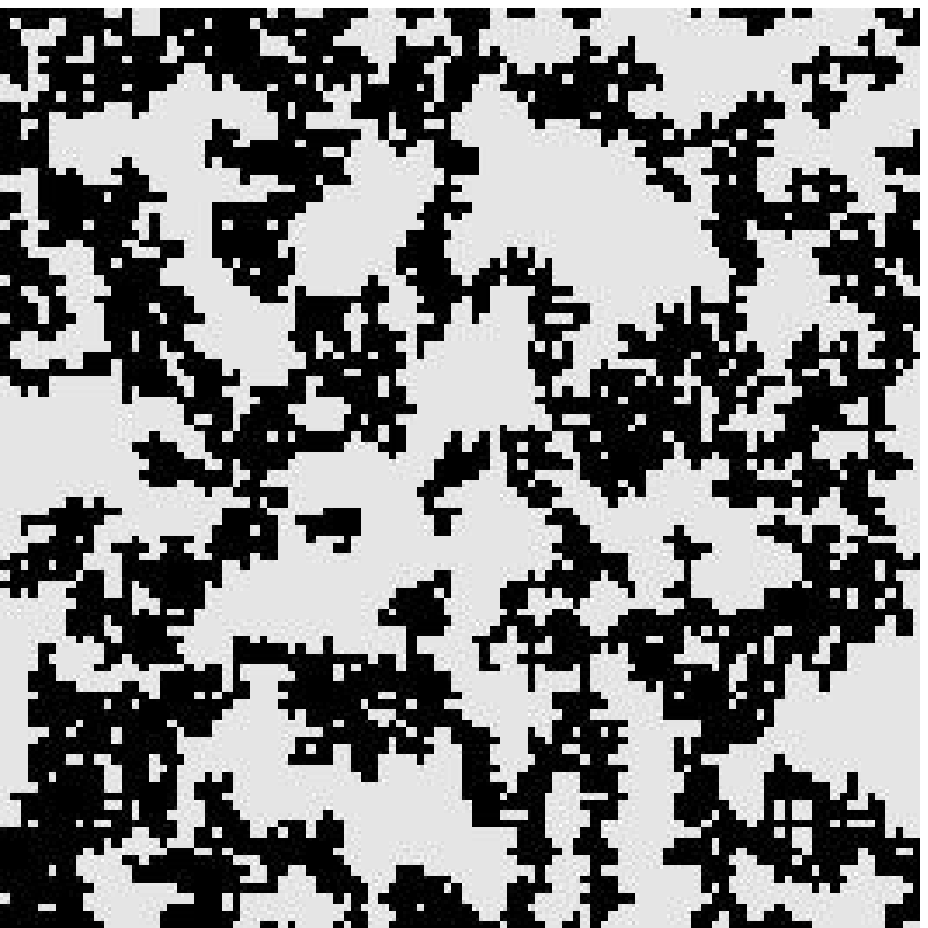}
\hfill
\includegraphics[width=2.5cm]{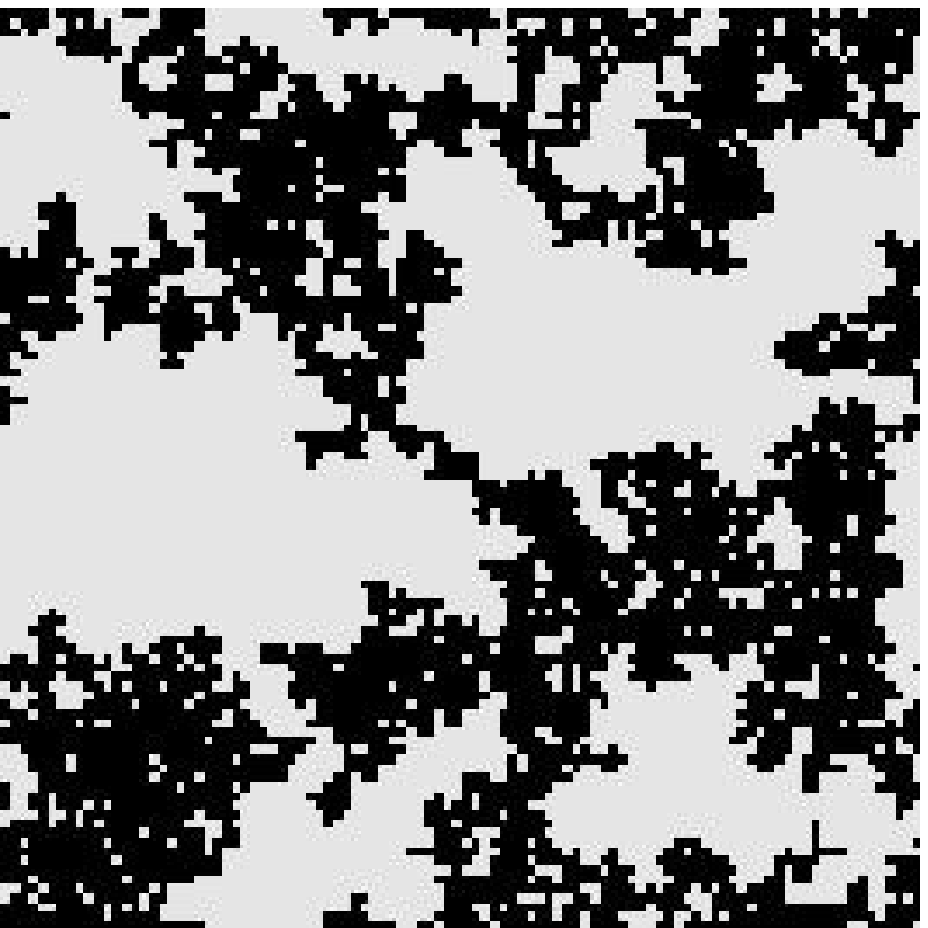}
\caption{\sl Direct visualization of dynamical heterogeneities with the
variables $\phi_i^*$,
for $\overline{\rho}=1.50, 1.65, 1.75$, in a system of size $N=100^2$
($\mu=1$). Only black sites ($\phi_i^*=0$) have changed state between $t=0$ and 
$t=\tau^*$. The typical size of both types of regions clearly increases with density.
}
\label{fig-Visu}
\end{figure}

A more detailed description of the heterogeneities is provided by the 
following (four-point) correlation:
\be
G_4(r,t) = [ \langle \phi_i(t) \phi_{i+r}(t) \rangle -
\langle \phi_i(t) \rangle \langle \phi_{i+r}(t) \rangle ], 
\ee
or its Fourier transform $\tilde{G}_4(k,t)$. One expects the rescaled correlation
$\overline{G}_4(k,t) \equiv \tilde{G}_4(k,t)/(t^{\nu} G_4(r=0,t))$
to scale as a function of $k\,t^{\nu}$ \cite{chi4}.
The corresponding data are plotted on Fig.~\ref{fig-subdif}(c), showing a
reasonable collapse.
Integrating $G_4(r,t)$ over $r$ yields the dynamical susceptibility
$\chi_4(t)$, which encodes some important information on the dynamics of the system \cite{chi4}.
Fig.~\ref{fig-subdif}(d) displays $\chi_4(t)/\chi_4^*$ versus $t/\tau^*$
for different $\overline{\rho}$, where
$\chi_4^*$ is the maximum of $\chi_4(t)$, and $\tau^*$ has been determined
from $\Phi(t)$. The resulting collapse is rather good,
although very long times would be needed to test quality of the collapse for $t \gg \tau^*$.
At short time ($t \ll \tau^*$), the data converges very slowly when $\overline{\rho} \to 2$ 
to the predicted power law $\chi_4(t) \sim t^{2\nu}$ \cite{chi4}, indicating strong sub-leading corrections.

Let us briefly discuss the qualitative behavior of the model in two
dimensions --a fuller account will be given in Ref.~\cite{long}.
It is interesting to visualize the variables $\phi_i^*$ for a given realization of the dynamics.
This is done on Fig.~\ref{fig-Visu}, for densities $\overline{\rho}=1.50$,
$1.65$ and $1.75$.  The typical size of the fast and slow regions is clearly seen to increase with $\overline{\rho}$.
A strong asymmetry appears: slow regions are essentially compact,
whereas fast regions seem to develop a fractal structure when their size
increases. Numerical results (not shown) indicate that in dimension $d=2$, the
cumulant $B(\overline{\rho},L)$ cannot be simply rescaled by the typical distance
$\eta^{-1/2}$ between mobility defects. Instead, an approximate rescaling
can be obtained using $\ell \sim \eta^{-\alpha}$, with $\alpha \approx 0.35$.
Mobility excitations exhibit a subdiffusive regime for $\mu>1$ and short times, 
before crossing over to pure diffusion. Such a cross-over is not observed in $d=1$, where subdiffusion seems
to hold at all times. 

In summary, we have analyzed a new class of {\sc kcm}s with a conserved, continuous density field, 
and a parameter $\mu$ accounting for interaction between particles. Although non trivial, 
the statics of the model can be worked out exactly. Our main result is that the motion of mobility excitations 
is subdiffusive for a large range of $\mu$, due to the presence of self-induced disorder.
The subdiffusion exponent varies continuously with $\mu$,
thus ruling out the possibility to describe the glassy dynamics in this model in terms of
standard directed percolation. An analytical understanding of the subdiffusive process, for instance
through a renormalization approach \cite{RG}, would be highly desirable.
Besides, experimental investigations of the (sub)diffusion
of mobility would also be
of great interest, and systems like sheared granular cells,
where subdiffusive motion of tracer particles has been
reported \cite{Marty}, may be promising candidates for such studies.

\end{document}